# A Comparative Analysis of the COVID-19 Infodemic in English and Chinese: Insights from Social Media Textual Data

Jia Luo[a], Daiyun Peng[a], Lei Shi [b,*], Didier El-baz [c], Xinran Liu[a]

a College of Economics and Management, Beijing University of Technology, Beijing, 100124, China.
E-mail: jialuo@bjut.edu.cn; daiyun_peng@outlook.com; liuxinranpro@outlook.com
b Guangxi Key Laboratory of Trusted Software, Guilin University of Electronic Technology, Guilin, 541004, China.
E-mail: leikyshi@88.com
c LAAS-CNRS, Université de Toulouse, CNRS, Toulouse, 31400, France.
E-mail: didier.el-baz@laas.fr
* Corresponding Author, E-mail: leikyshi@88.com

## Abstract

The COVID-19 infodemic, characterized by the rapid spread of misinformation and unverified claims related to the pandemic, presents a significant challenge. This paper presents a comparative analysis of the COVID-19 infodemic in the English and Chinese languages, utilizing textual data extracted from social media platforms. To ensure a balanced representation, two infodemic datasets were created by augmenting previously collected social media textual data. Through word frequency analysis, the thirty-five most frequently occurring infodemic words are identified, shedding light on prevalent discussions surrounding the infodemic. Moreover, topic clustering analysis uncovers thematic structures and provides a deeper understanding of primary topics within each language context. Additionally, sentiment analysis enables comprehension of the emotional tone associated with COVID-19 information on social media platforms in English and Chinese. This research contributes to a better understanding of the COVID-19 infodemic phenomenon and can guide the development of strategies to combat misinformation during public health crises across different languages.

## Keywords

COVID-19; Infodemic Data; Word Frequency Analysis; Topic Clustering Analysis; Sentiment analysis

### 1 Introduction

During the beginning of the COVID-19 pandemic, there was a surge in misinformation, false information, and rumors spreading rapidly across various media platforms. This phenomenon came to be known as the infodemic. The infodemic refers to the overwhelming abundance and rapid spread of misinformation, conspiracy theories, and unverified claims related to the pandemic [1]. It accompanied the spread of the virus itself and was fueled by uncertainties, fear, and confusion during the early stages of the outbreak [2]. Numerous falsehoods and conspiracy theories circulated globally, making it challenging for individuals to discern accurate information. Some examples include claims that the virus was intentionally created or released, that certain medications or alternative remedies could cure or prevent the virus, or that 5G networks were somehow linked to the spread of the disease. The infodemic had significant implications on public health, as it hindered effective pandemic response efforts. False information about prevention measures, symptoms, and treatments could potentially mislead the public and endanger lives. It also led to widespread panic, social unrest, and stigmatization of certain groups. Although the World Health Organization (WHO) declared an end to COVID-19 as a global health emergency, it is important to note that combating the infodemic remains an ongoing challenge.

COVID-19 is a global pandemic, and misinformation knows no boundaries. Conducting a comparative analysis across different languages allows us to gain a comprehensive, global perspective on the infodemic. On one side, each language has its unique linguistic characteristics, cultural norms, and online behaviors. Analyzing social media data in different languages can uncover language-specific nuances that shape misinformation patterns and responses. On the other side, analyzing data from multiple languages uncovers common themes, misinformation tactics, and influential narratives. Understanding these cross-cultural trends allows for exchanging knowledge and implementing effective global strategies. According to Statista [3], English and Chinese are the top-2 most common languages used on the Internet. Therefore, this paper aims to conduct a comparative analysis of the COVID-19 infodemic in both English and Chinese languages by utilizing textual data extracted from social media platforms. The main contributions of our work are summarized as follows:

1. Two balanced infodemic datasets are introduced by adjusting previously collected social media textual data with annotations from healthcare workers where all records are classified into three distinct groups: true, false, and uncertain.
2. Word frequency analysis is conducted to identify the thirty-five most frequently occurring infodemic words to acquire knowledge on the prevalent patterns and trends of word usage in two languages.
3. Topic clustering analysis is executed to uncover thematic structures to gain insights into the similarities and differences between different topics or subject areas across two languages.
4. Sentiment analysis is performed to determine the percentage of positive, neutral, or negative sentiments within infodemic records to understand the emotional tone and attitudes expressed in two languages.
5. A discussion is held to grasp the language-specific nuances and cross-cultural trends of both the overall records and the records classified into three groups. The latter offers perspectives at a more refined level by incorporating the professional knowledge of healthcare workers.

The subsequent sections of this paper are organized as follows. Section 2 introduces related works. Section 3 displays the two balanced infodemic datasets. Section 4 provides the results of word frequency analysis, topic clustering analysis, and sentiment analysis respectively. Afterward, a discussion is illustrated in Section 5. Finally, section 6 presents conclusions.

## 2 Related Works

The majority of scholarly research about infodemic centers on addressing misinformation while trained models incorporating word embeddings stand out as the most commonly utilized methods [4]. Glazkova et al. [5] proposed an approach using the transformer-based ensemble of COVID-Twitter-BERT models to detect COVID-19 fake news in English. Chen et al. [6] studied a novel transformer-based language model fine-tuning approach for English fake news detection during COVID-19. Paka et al. [7] set up a cross-stitch semi-supervised neural attention model for COVID-19 fake news detection which leverages the large amount of unlabelled data from Twitter in English. Chen et al. [8] used fuzzy theory to extract features and designed multiple deep-learning model frameworks to identify Chinese and English COVID-19 misinformation. Liu et al. [9] developed a deep learning base model and fine-tuned it to adapt to the specific domain context of COVID-19 news classification in English, Chinese, Arabic, and German. While these models have undoubtedly improved the efficacy in combatting misinformation during the COVID-19 pandemic, they often overlook the critical aspect of elucidating the underlying characteristics of the infodemic. Without being transformed into human-understandable knowledge, their outputs would have limited efficacy in aiding human efforts to combat the infodemic and develop targeted countermeasures and mitigation strategies.

Certain academic studies pay their attention to comprehending the patterns exhibited within the COVID infodemic through an in-depth analysis of its content. Gupta et al. [10] identified topics and key themes present in English COVID-19 fake and real news, compared the emotions associated with these records and gained an understanding of the network-oriented characteristics embedded within them. Wan et al. [11] described the prominent lexical and grammatical features of English COVID-19 misinformation, interpreted the underlying (psycho-)linguistic triggers, and studied the feature indexing for anti-infodemic modeling. Zhao et al. [12] used 1296 COVID-19 rumors collected from an online platform in China, and found measurable differences in the content characteristics between true and false rumors. Zhou et al. [13] investigated both thematic and emotional characteristics of COVID-19 fake news at different levels and compared them in English and Chinese. All of the aforementioned works prioritize conducting analysis using a binary truth classification system, precisely distinguishing between true and false categories, to minimize discrepancies arising from truth labeling. However, it is incumbent upon us to acknowledge the inherent challenges faced when adjudicating the authenticity or veracity of certain statements during the labeling process.

The majority of collected records utilized in the analysis and detection of the infodemic phenomenon are typically categorized and labeled as either true or false [14]. Nonetheless, a limited number of studies have undertaken an alternative approach by classifying these records into 3-5 categories to have a more comprehensive understanding of the infodemic and its impact at a finer level of granularity. Cheng et al. [15] built up an English COVID-19 rumor dataset by gathering news and tweets and manually labeling them as true, false, or unverified. Haouari et al. [16] proposed an Arabic COVID-19 Twitter dataset where each tweet was marked as true, false, or others. Luo et al. [17] collected widely spread Chinese infodemic during the COVID-19 outbreak from Weibo and WeChat while each record was indicated as true, false, or questionable after a four-time adjustment. Kim et al. [18] produced a dataset encompassing English claims and corresponding tweets, which were organized into four groups: COVID true, COVID fake, non-COVID true, and non-COVID fake. Dharawat et al. [19] released a dataset for health risk assessment of COVID-19-related social media posts. There are English tweets and tokens and all of them were classified into five categories: real news/claims, not severe, possibly severe misinformation, highly severe misinformation, or refutes/rebuts misinformation. Given the profound interconnectedness between the infodemic and health records and its notable implications for public health, the active involvement of healthcare workers could help advance the comprehension of the infodemic. However, only [17] have considered this aspect while categorizing the collected records.

Considering the above-mentioned analysis, most studies have predominantly focused on English records. Therefore, it is valuable to conduct a comparative study of the COVID-19 Infodemic in multiple languages. The previously collected social media textual data offer an initial starting point while the integration of healthcare workers' professional knowledge serves to enhance insights at a more refined level. Additionally, conducting an analysis incorporating lexical, topical, and sentiment features would contribute to a comprehensive understanding of the underlying characteristics.

## 3 Data Collection

English and Chinese records are chosen for this study because of their status as the two most prevalent languages used on the Internet [3]. A summary of the encompassed data is presented in Table 1.

The English data is sourced from [20]. It collects 5100 fake news from public fact-verification websites and social media. On the other side, there are 5600 real news and they are tweets crawled from official and verified Twitter handles of the relevant sources using Twitter API. The dataset is split into train (60%), validation (20%), test (20%) and the training set has been selected for this study. The training set was published on October 1, 2020 [21] and consists of 3360 real news and 3060 fake news. We have invited three healthcare workers to manually classify these 6420 records into three distinct groups: true, false, and uncertain. Their assessments rely exclusively on their judgments without any reference to external sources, and the assigned label for each record is determined by employing a majority agreement methodology. To address the limited number of instances in the real group (830 records), we randomly selected 830 records from both instances labeled as true and uncertain. Finally, a total of 2490 records were kept, with an equal distribution for each group to mitigate any potential bias and to ensure fairness in representing various categories.

The Chinese data is derived from [17]. This dataset gathers a total of 797 original records, which include manually verified Weibo posts from the Sina Community Management Center between January 21 and April 10, 2020, and specifically checked news from the WeChat mini-program "Jiaozhen" until March 31, 2020. All instances are classified into two types based on their content: strongly related health records and weakly related health records. The weakly related health records are further subdivided into specific categories, which include local measures, national measures, patient information, and others. Subsequently, four rounds of adjustments are conducted: (1) adjusting labels for instances classified as weakly related health records, (2) adjusting labels for records initially marked as partially true or conditionally true, (3) removing dummy records in the sub-group of local measures, (4) adding strongly related health records from authoritative sources to the true group. In the end, the dataset consists of 1055 records overall, with 409 labeled as questionable, 276 as false, and 335 as true, ensuring that each group contains roughly an equal number of records. Since there is high intercoder reliability between the final labels and labels annotated by healthcare workers, we keep the classification results from [17] while simply replacing the label questionable with uncertain.

Table 1 A summary of the encompassed data

| Languages | Sources | Labels | | |
|---|---|---|---|---|
| | | True | False | Uncertain |
| English | Patwa et al. [20] | 830 | 830 | 830 |
| Chinese | Luo et al. [17] | 335 | 276 | 409 |

## 4 Methods and Results

### 4.1 Word Frequency Analysis

Weiciyun [22] is utilized in this section to conduct word frequency analysis for both English and Chinese records. It serves as a practical and user-friendly online tool for generating word clouds and visualizing text data. Before analysis, the built-in language-specific tokenization and stopword removal techniques provided by Weiciyun are leveraged to yield clean and meaningful text data. Afterward, content filtration based on part-of-speech is applied to retain only nouns, gerunds, and proper nouns. In terms of English text, only content with a word length of at least 3 and a frequency of at least 2 is selected. Similarly, for Chinese text, content with a character length of at least 2 is chosen. Finally, the thirty-five most frequent words are presented and they are illustrated with font size scaled to their frequencies while the detailed word frequencies of these words can be found in Table 2. To ensure translation consistency and reduce subjectivity, the word clouds maintain the original Chinese characters while providing a reference translation in Appendix 1 as needed.

Table 2 Word frequency of the thirty-five most frequent words displayed in Figure 1 and Figure 2 (W.F. = Word Frequency)

| All records | | | | Records labeled as true | | | | Records labeled as false | | | | Records labeled uncertain | | | |
|---|---|---|---|---|---|---|---|---|---|---|---|---|---|---|---|
| English | W.F. | Chinese | W.F. | English | W.F. | Chinese | W.F. | English | W.F. | Chinese | W.F. | English | W.F. | Chinese | W.F. |
| Covid | 833 | 病毒 | 185 | Covid | 476 | 病毒 | 71 | Coronavirus | 353 | 病毒 | 71 | Cases | 336 | 肺炎 | 84 |
| Coronavirus | 617 | 肺炎 | 179 | People | 141 | 口罩 | 60 | People | 91 | 肺炎 | 59 | Covid | 274 | 武汉 | 48 |
| Cases | 430 | 口罩 | 110 | Spread | 126 | 肺炎 | 36 | Covid | 83 | 口罩 | 32 | Coronavirus | 162 | 病毒 | 43 |
| People | 319 | 疫情 | 56 | Coronavirus | 102 | 患者 | 28 | Virus | 79 | 疫情 | 13 | Tests | 123 | 疫情 | 39 |
| Health | 177 | 武汉 | 55 | Health | 97 | 消毒剂 | 22 | Trump | 62 | 患者 | 11 | Deaths | 103 | 医院 | 20 |
| Tests | 159 | 患者 | 54 | Risk | 87 | 症状 | 17 | Pademic | 51 | 美国 | 10 | Number | 101 | 美国 | 20 |
| Spread | 147 | 美国 | 30 | Cases | 76 | 医用 | 16 | Cure | 51 | 酒精 | 10 | People | 87 | 中国 | 19 |
| Deaths | 145 | 钟南山 | 26 | Face | 73 | 飞沫 | 14 | President | 49 | 钟南山 | 9 | States | 85 | 口罩 | 18 |
| Virus | 138 | 消毒剂 | 26 | Others | 67 | 建议 | 11 | Vaccine | 47 | 疫苗 | 8 | India | 78 | 钟南山 | 16 |
| Testing | 137 | 酒精 | 25 | Testing | 63 | 风险 | 10 | Video | 42 | 武汉 | 7 | Today | 71 | 患者 | 15 |
| Pademic | 125 | 疫苗 | 24 | Symptoms | 62 | 酒精 | 10 | Government | 38 | 大蒜 | 6 | Testing | 64 | 北京 | 15 |
| Vaccine | 119 | 医院 | 22 | Patinets | 52 | 疾病 | 10 | Corona | 37 | 大量 | 6 | State | 62 | 上海 | 14 |
| Number | 119 | 中国 | 22 | Virus | 50 | 证据 | 10 | China | 37 | 病人 | 5 | Indiafightscorona | 59 | 意大利 | 14 |
| States | 116 | 病人 | 20 | Pandemic | 47 | 感染者 | 9 | News | 33 | 日本 | 5 | Health | 47 | 病人 | 11 |
| India | 111 | 症状 | 20 | Masks | 46 | 人群 | 9 | Health | 33 | 抗体 | 4 | Report | 44 | 疫苗 | 10 |
| Patients | 109 | 医用 | 18 | Mask | 46 | 儿童 | 8 | Claims | 32 | 院士 | 4 | Vaccine | 43 | 病例 | 9 |
| Risk | 107 | 风险 | 17 | Care | 44 | 居家 | 8 | Chinese | 31 | 医生 | 4 | Rate | 42 | 湖北 | 9 |
| Face | 87 | 北京 | 17 | Hands | 43 | 通风 | 8 | Masks | 31 | 病毒感染 | 4 | Case | 41 | 人员 | 9 |
| Test | 87 | 人员 | 17 | Use | 42 | 人员 | 8 | Bill | 28 | 空气 | 4 | Nigeria | 39 | 成都 | 9 |
| Trump | 84 | 病例 | 16 | Contact | 42 | 物品 | 7 | World | 28 | 白酒 | 3 | Data | 38 | 院士 | 8 |
| State | 83 | 意大利 | 16 | Distancing | 40 | 效果 | 7 | Gates | 27 | 防病毒 | 3 | Lakh | 38 | 入境 | 7 |
| Masks | 83 | 建议 | 15 | Home | 39 | 传染性 | 7 | Flu | 26 | 小时 | 3 | Lockdown | 37 | 医生 | 7 |
| Days | 82 | 上海 | 14 | CDC | 39 | 人类 | 7 | Novel | 25 | 病情 | 3 | Day | 35 | 视频 | 7 |
| Today | 81 | 飞沫 | 14 | Measures | 37 | 距离 | 7 | Donald | 25 | 中国 | 3 | Patients | 35 | 全国 | 7 |
| Symptoms | 81 | 抗体 | 13 | Cloth | 35 | 核酸检测 | 6 | Being | 24 | 流鼻涕 | 3 | Days | 32 | 全部 | 6 |
| Indiafightscorona | 75 | 院士 | 13 | Disease | 34 | 疫苗 | 6 | India | 24 | 纸尿裤 | 3 | Test | 32 | 阳性 | 6 |
| Others | 73 | 感染者 | 13 | Test | 34 | 动物 | 6 | Claim | 23 | 气溶胶 | 3 | Yesterday | 31 | 员工 | 6 |
| Home | 72 | 阳性 | 12 | Treatment | 30 | 食品 | 6 | Outbreak | 22 | 二氧化氯 | 3 | Week | 31 | 印度 | 6 |
| CDC | 71 | 核酸检测 | 12 | Days | 29 | 情况 | 6 | Home | 22 | 消毒剂 | 3 | First | 30 | 国家 | 5 |
| Government | 71 | 医生 | 11 | Vaccine | 29 | 传播者 | 6 | Lockdown | 22 | 牛羊肉 | 3 | Million | 30 | 物资 | 5 |
| Data | 70 | 疾病 | 11 | Data | 29 | 重症 | 6 | Patients | 22 | 喉咙 | 3 | Recoveries | 29 | 酒精 | 5 |
| Lockdown | 70 | 湖北 | 11 | Infection | 29 | 手部 | 6 | Disease | 21 | 肥皂 | 3 | Coronavirusupdates | 28 | 特朗普 | 5 |
| Care | 69 | 空气 | 11 | Countries | 29 | 手套 | 6 | Test | 21 | 食品 | 3 | Pandemic | 27 | 风险 | 5 |
| Video | 68 | 证据 | 11 | Deaths | 27 | 传染病 | 6 | Days | 21 | 食用 | 3 | Isolation | 27 | 广州 | 5 |
| Case | 67 | 人类 | 10 | Person | 27 | 紫外线 | 5 | Message | 20 | 瘟疫 | 2 | Numbers | 27 | 医疗 | 5 |

The word clouds of all records in English and Chinese are presented in Figure 1. Firstly, it is noteworthy that the most frequently mentioned terms in both languages are the same, including "virus" (病毒), "pandemic" (疫情), "patient" (患者), and so on. Secondly, the term "mask" (口罩) is mentioned in both languages but holds greater prominence in the Chinese word cloud. Thirdly, the name "Wuhan" (武汉), which corresponds to the initial epicenter of the COVID-19 outbreak in China, appears in larger font size in the Chinese word cloud, while no specific city-related word is present in the English cloud. Fourthly, the term "death" appears with greater frequency in the English data than in the Chinese records where it is noticeably absent. Finally, the individual most frequently mentioned in English is President Donald Trump, whereas, in Chinese, it is Zhong Nanshan (钟南山), an esteemed academician in the field of healthcare.

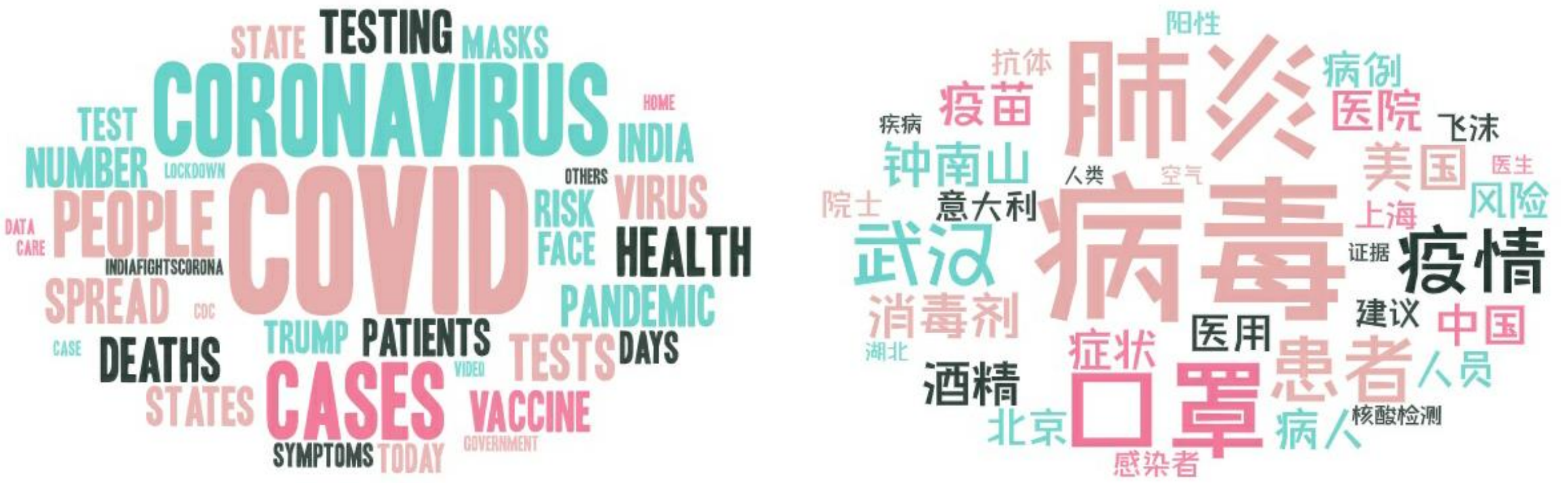

Figure 1. Word clouds for all records

The word clouds presented in Figure 2 categorize records into three groups in both English and Chinese. The true or false labeled groups primarily consist of common terms, which are predominantly derived from the expertise of healthcare professionals. These terms revolve around virus transmission methods, prevention measures, and treatment approaches. On the other hand, the uncertain group encompasses a diverse range of terms. Within this group, both English and Chinese records demonstrate an awareness of regional considerations. Notably, the Chinese word cloud places a greater emphasis on specific locations such as "Wuhan" (武汉), "Beijing" (北京), "Shanghai" (上海), "Canton" (广州), and "Chengdu" (成都). In contrast, the English word cloud labeled as uncertain indicates a temporal focus by frequently including terms like "Today," "Yesterday," "Days," and "Week." It is worth mentioning that these time-related terms are not explicitly included in the Chinese word cloud.

Word clouds of records labeled as true

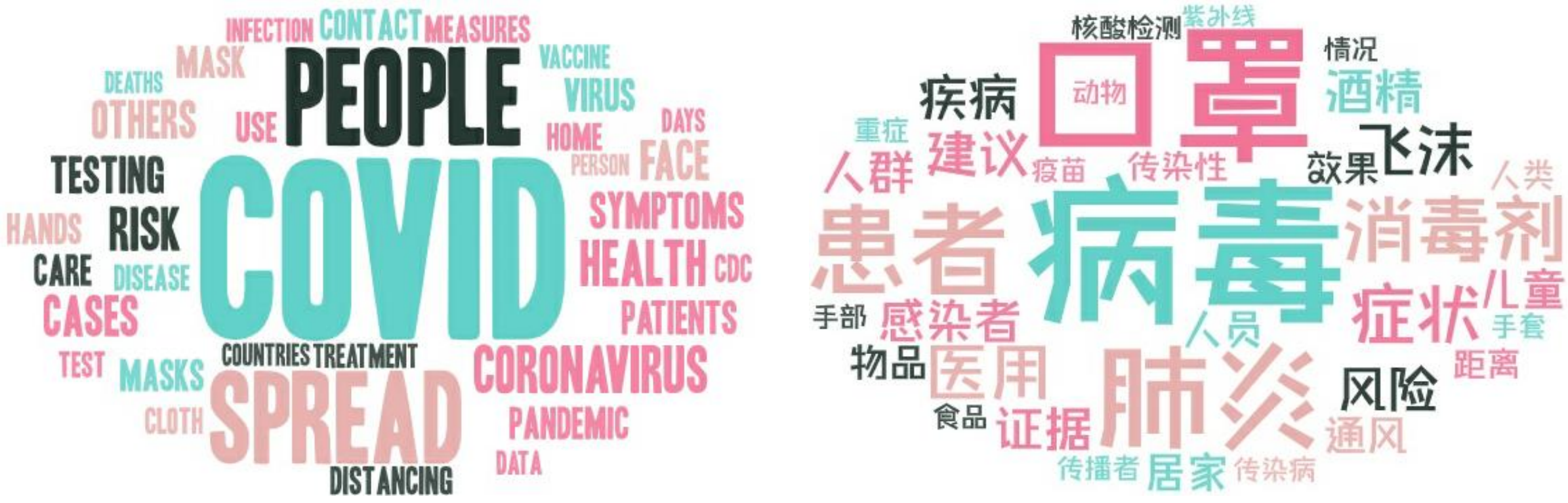

Word clouds of records labeled as false

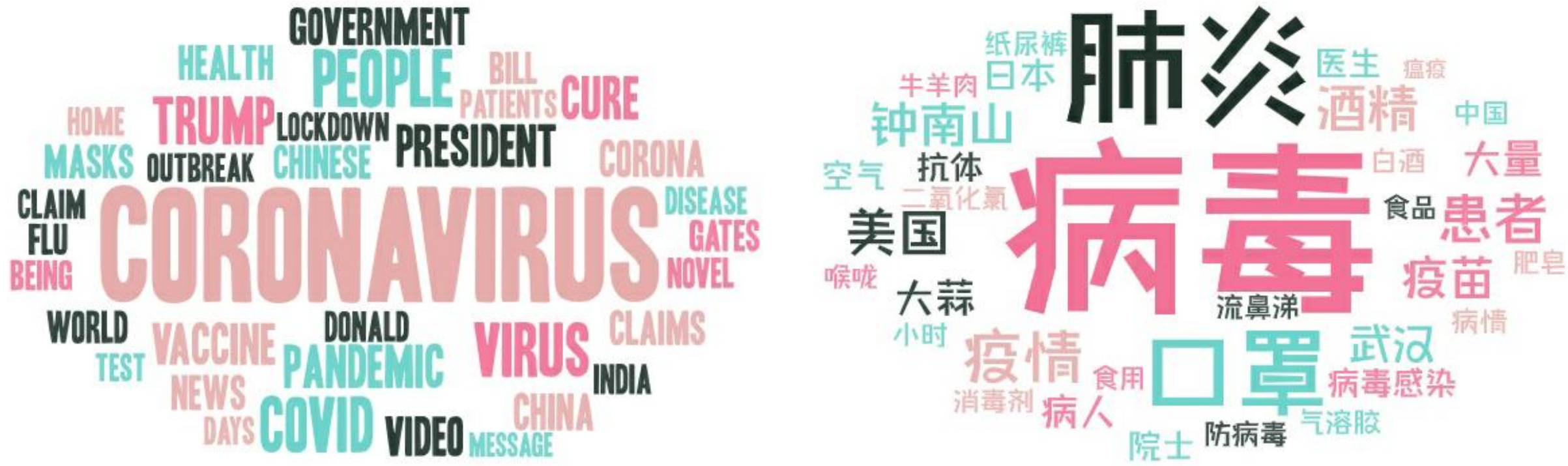

Word clouds of records labeled as uncertain

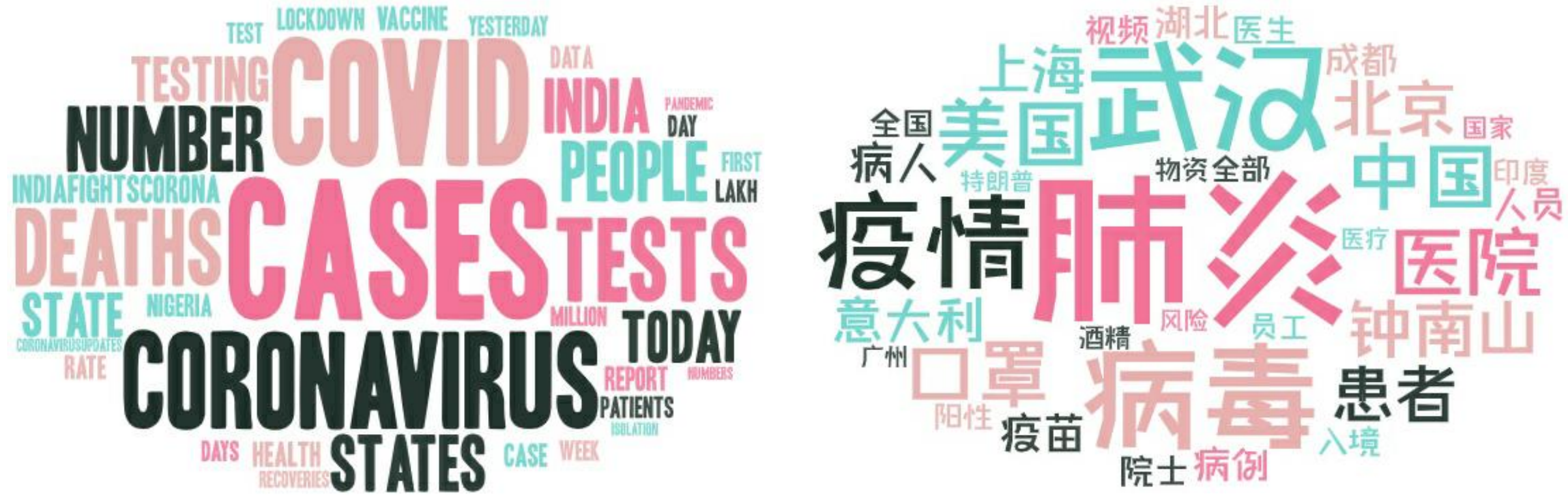

Figure 2. Word clouds for records classified into three groups

### 4.2 Topic Clustering Analysis

In this section, the Latent Dirichlet Allocation (LDA) topic model is implemented to uncover hidden topics and thematic structures from both English and Chinese records. LDA is a widely adopted technique in the field of natural language processing, wherein documents are represented as stochastic mixtures across latent topics, and each topic is characterized by a distribution over words [23]. For enhanced comprehension of the clustered topics, we employ the LDAvis package [24] to visualize the results using multidimensional scale analysis. We set the initial range of topic numbers to [1, 15], and the final determination of the optimal number of topics relies on the highest coherence score. The step size is retained as 1, while $\alpha$ and $\beta$ are maintained at their default values. Furthermore, language-specific tokenization and stopword removal techniques are employed to mitigate the influence of text analysis when applying LDA to analyze different languages. For English text, whitespace-based tokenization is employed, and the widely recognized Chinese word segmentation tool Jieba is utilized for Chinese tokenization. The built-in function from the Natural Language Toolkit (NLTK) library in Python is leveraged to access a collection of stopwords specifically for English, whereas the widely used cn_stopwords.txt file is applied to remove stopwords from Chinese text. Finally, in line with sub-section 4.1, the original Chinese characters are preserved in the visualization graphs, supplemented with a reference translation provided in Appendix 2.

The visualization graphs of all records in English and Chinese are presented in Figure 3. Firstly, the number of clustered topics in the English records is significantly fewer compared to the Chinese records. Specifically, there are only 4 topics identified in the English records, whereas the Chinese records encompass 13 topics. Secondly, the English topics are mutually exclusive with no overlap. The proximity between Topic 1 and Topic 2 is high, while the remaining topics exhibit considerable dissimilarity. Conversely, in the visualization graph of the Chinese records, the topics demonstrate interconnectedness. Notably, Topic 2 overlaps with Topic 9, as does Topic 8 with Topic 11. Thirdly, Topic 1 stands out in the English records as it covers a significant portion of the tokens, specifically 35.2% in the top 30 most relevant terms. On the other hand, Topic 1 has a comparatively smaller presence in the Chinese records, accounting for only 11% of the tokens in the top 30 most relevant terms. Its size is not as noticeable when compared to Topic 2 and Topic 3, where the difference is not considered significant. Finally, there are shared terms that appear in the top 30 most relevant terms of Topic 1 in both languages, indicating a mutual focus from both sides.

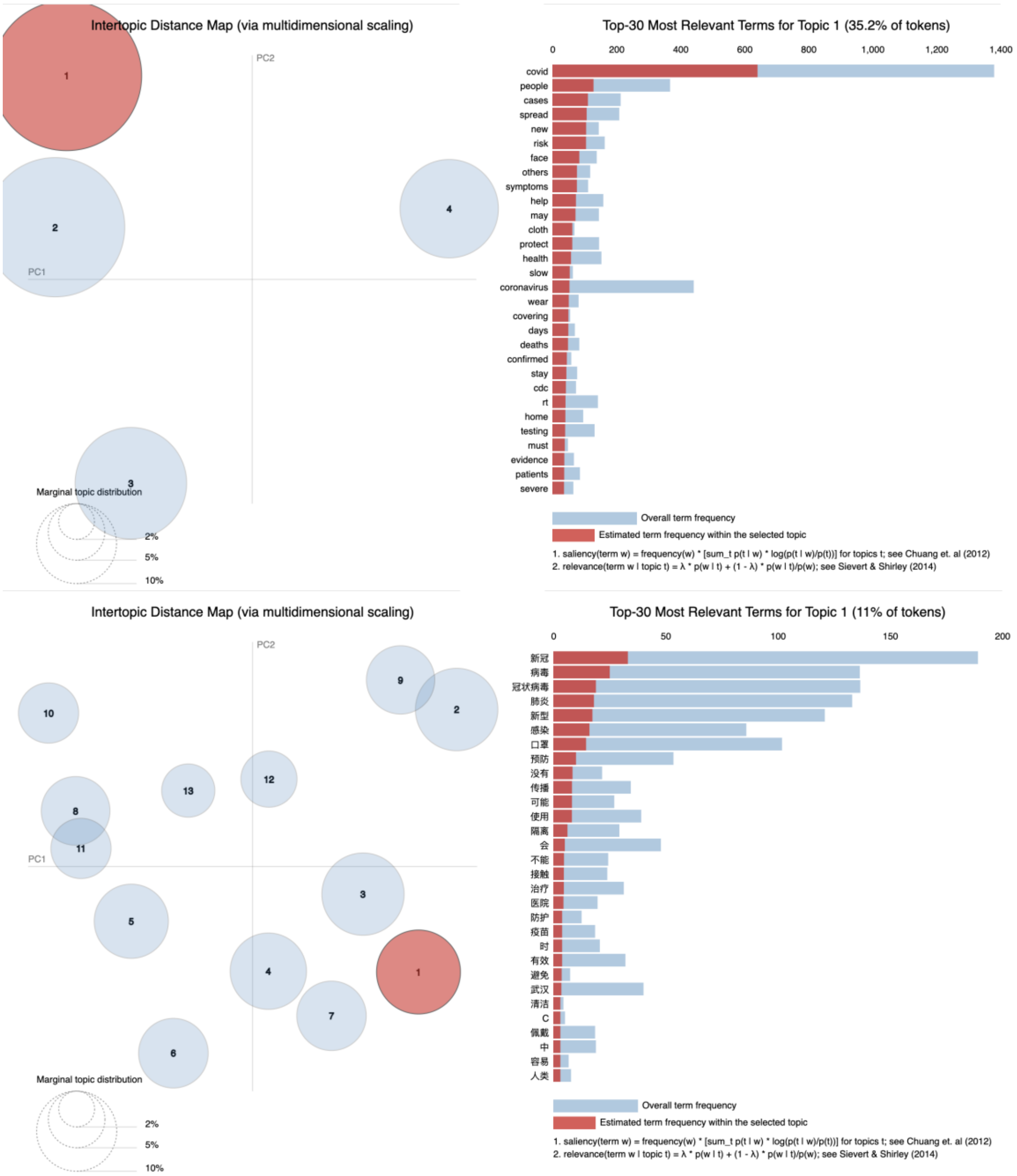

Figure 3. Visualization graph of LDA topic modeling for all records

Visualization graph of LDA topic modeling for records labeled as true

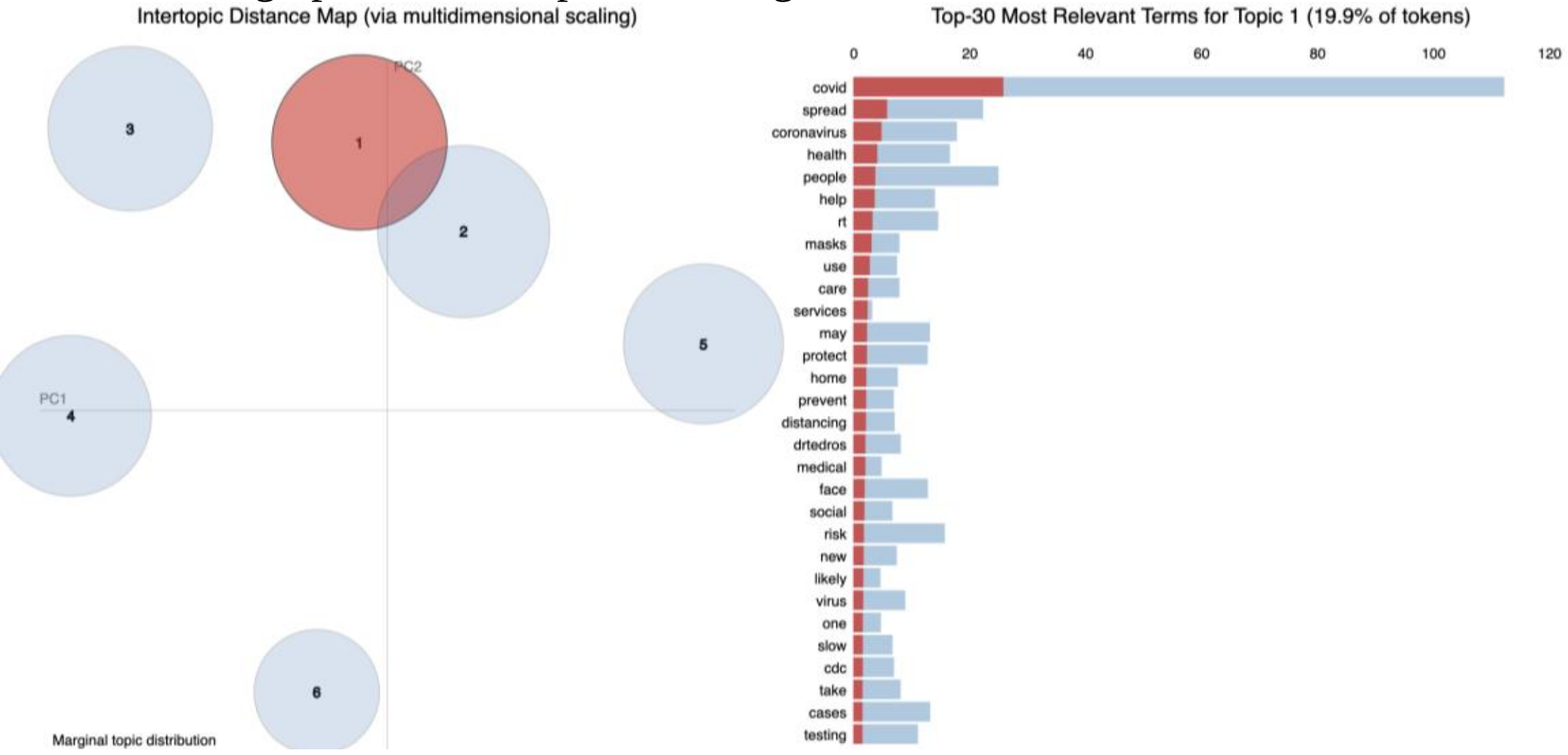

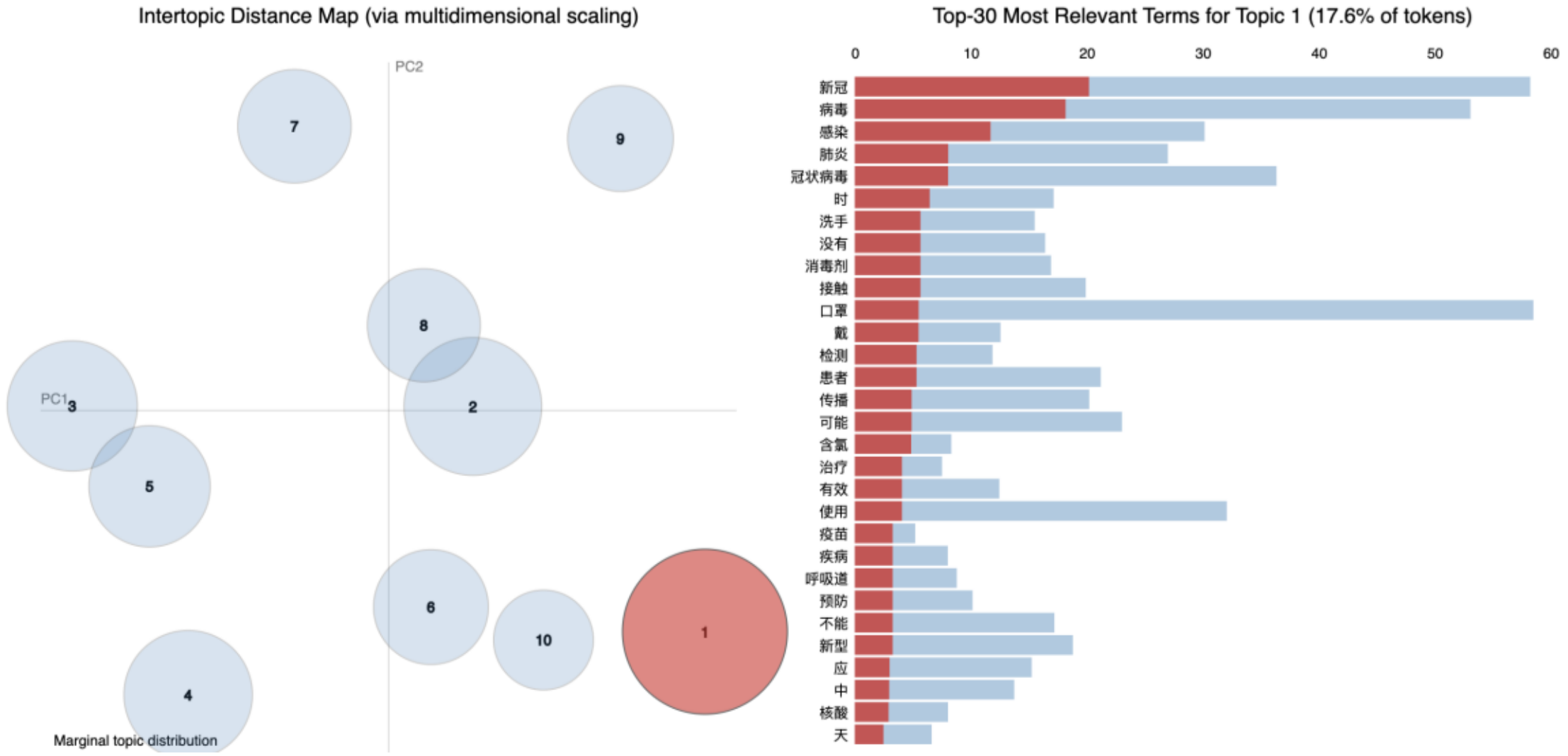

Visualization graph of LDA topic modeling for records labeled as false

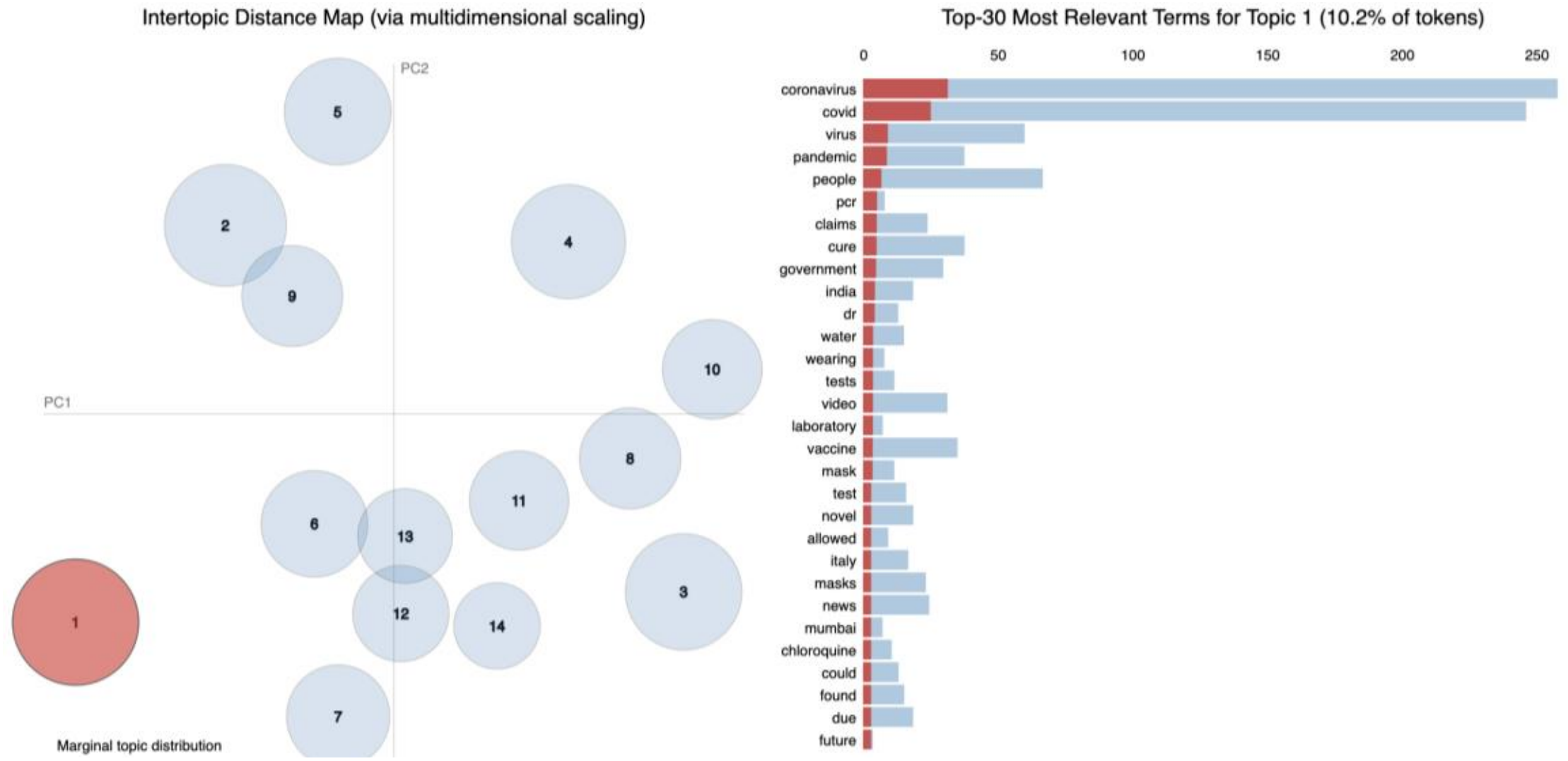

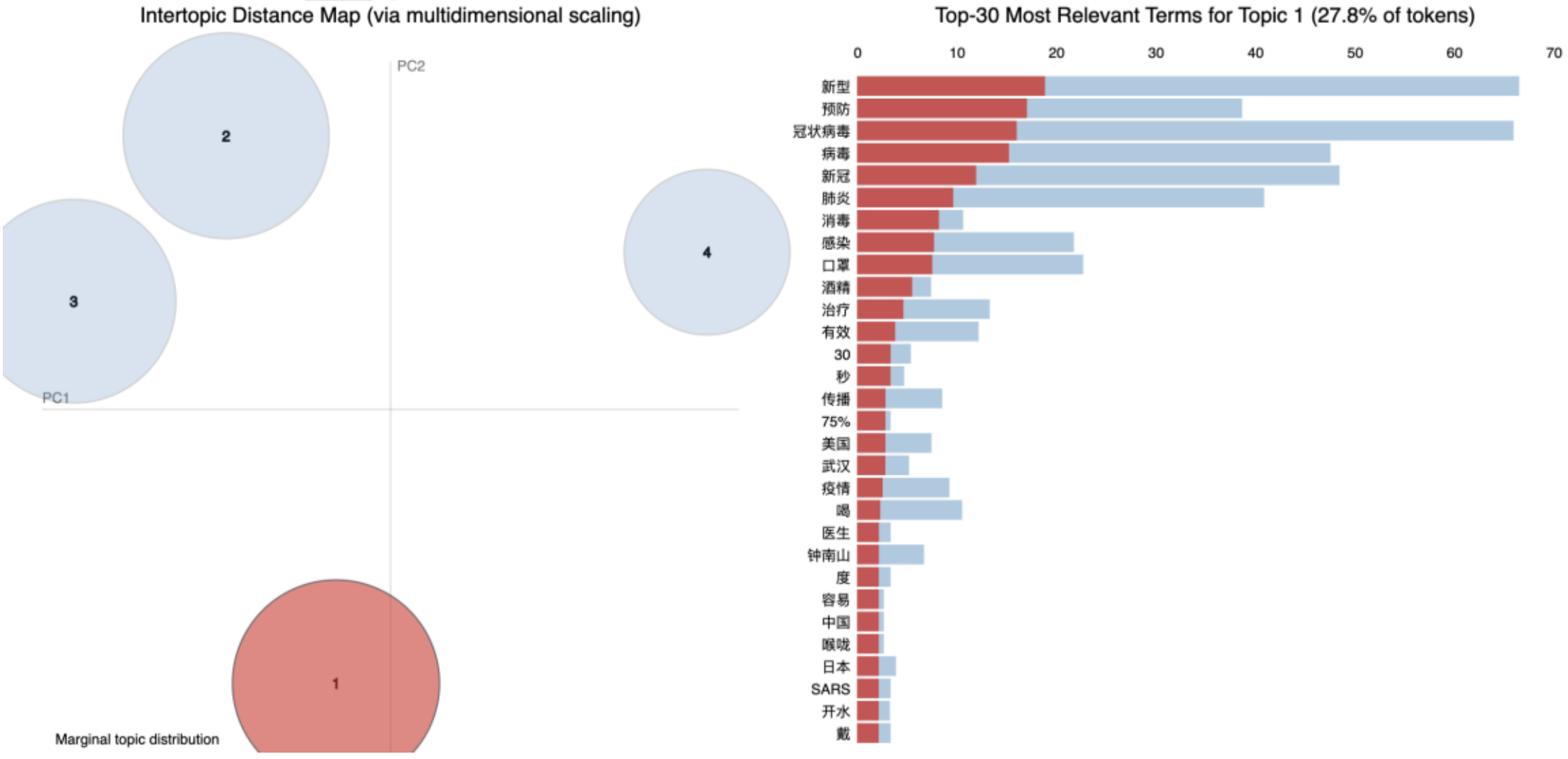

Visualization graph of LDA topic modeling for records labeled as uncertain

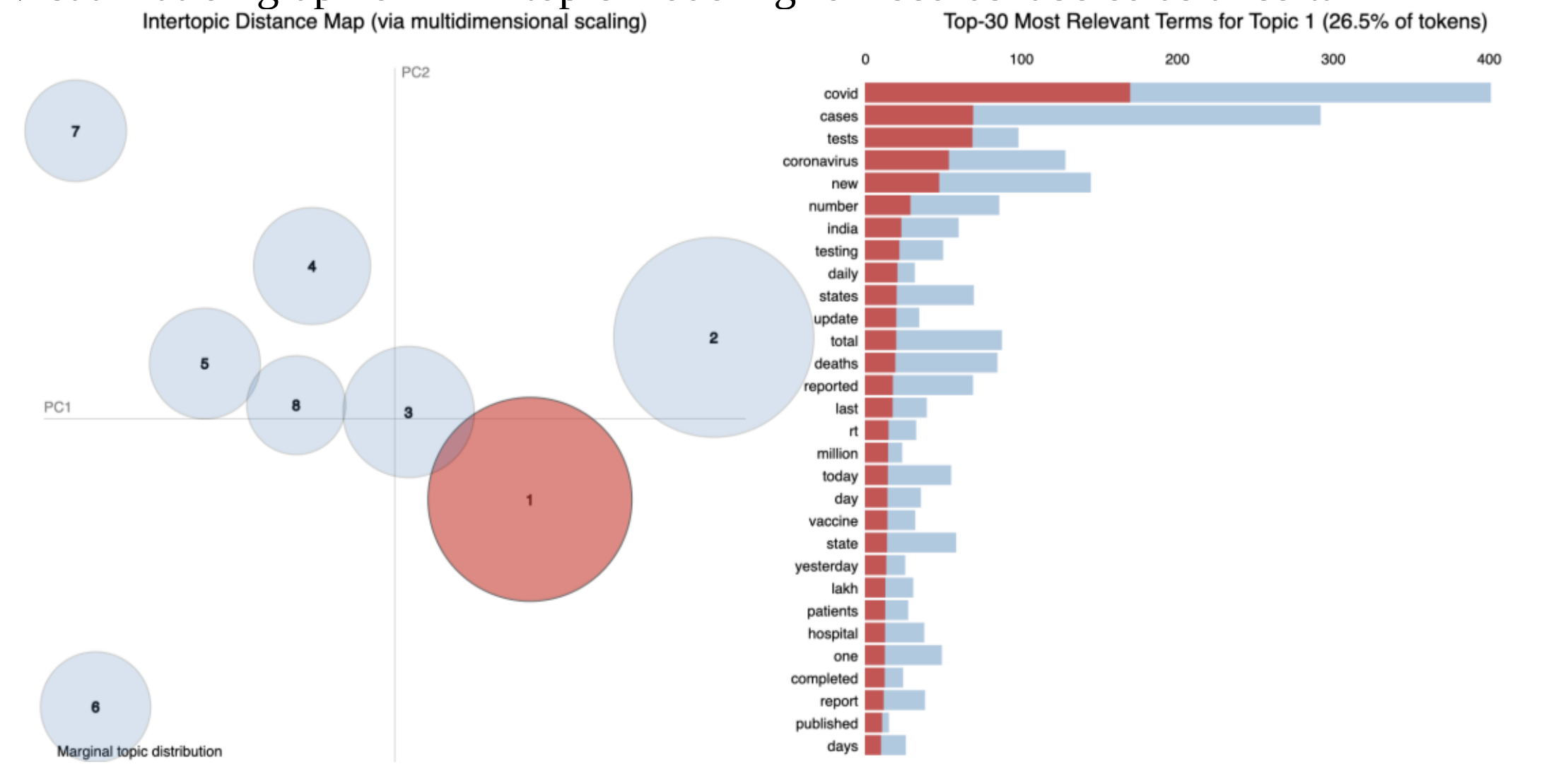

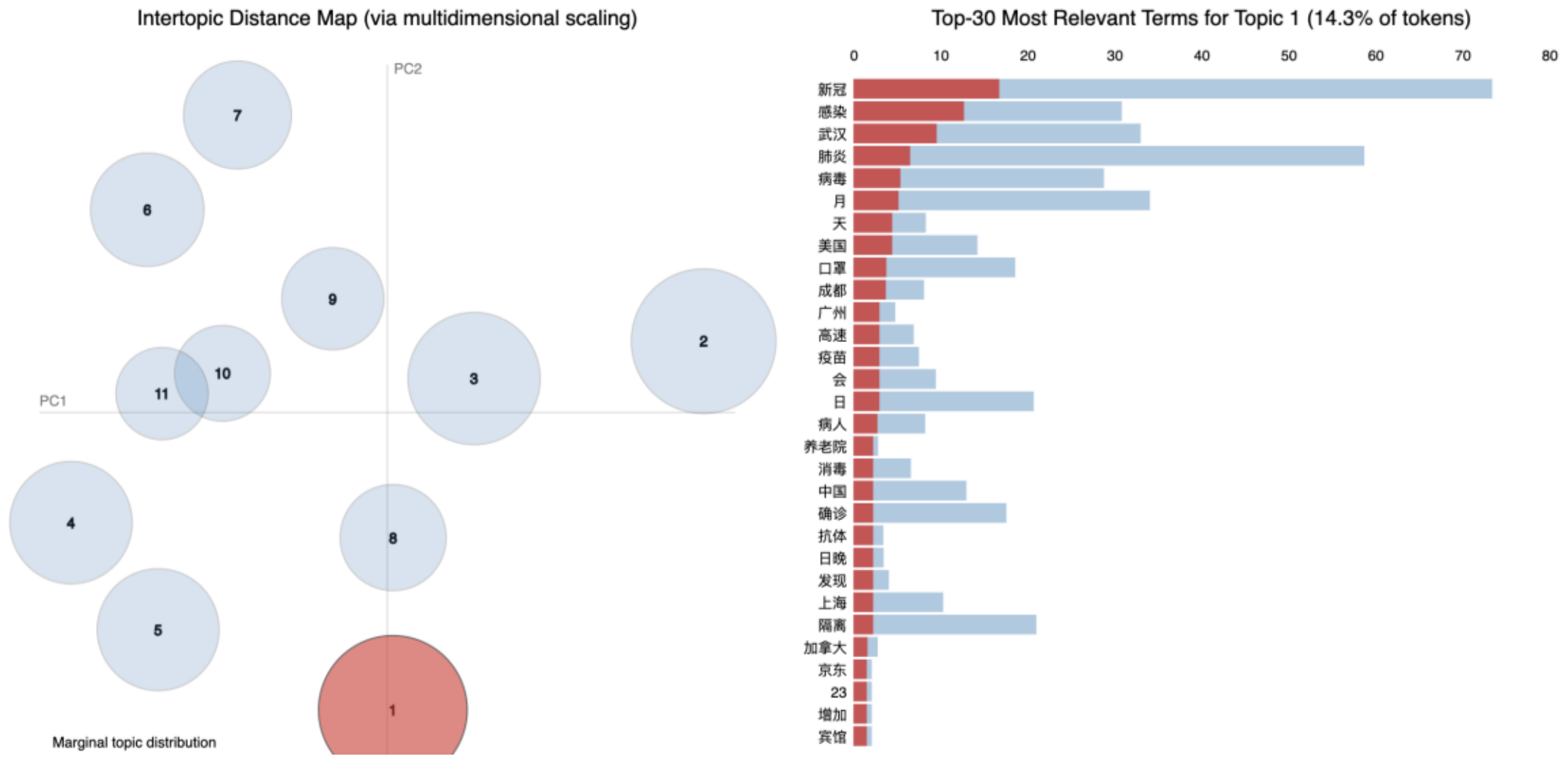

Figure 4. Visualization graph of LDA topic modeling for records classified into three groups

The visualization graphs in Figure 4 categorize records into three groups in both English and Chinese. The annotation of each visualization graph remains the same as shown in Figure 3. Due to space limitations, they are not included in Figure 4. Firstly, the pattern of topic numbers remains consistent across the groups labeled as true and uncertain. However, in the group labeled as false, the English records show a significantly larger number compared to the Chinese records. Secondly, within the groups labeled as true, the percentage of tokens in the top 30 most relevant terms of Topic 1 is similar in both languages while there exists a difference of more than 10% in the other two groups. Finally, the groups labeled as true or false primarily consist of common terms in the top 30 most relevant terms in both languages. Nevertheless, the uncertain group encompasses a diverse range of terms. This observation further supports the conclusion mentioned in sub-section 4.1.

### 4.3 Sentiment analysis

Monkeylearn [25] is utilized in this section to conduct sentiment analysis on English records. The platform offers a user-friendly graphical interface that enables users to create personalized text classification and extraction analyses by training machine learning models. In the analysis of Chinese records, ROST_CM6 [26], a widely used Chinese social computing platform, is employed to generate the results. ROST_CM6 enables various text analyses, including microblog, chat, and web-wide analyses. It is important to note that Monkeylearn generated multiple emotions for 145 instances due to the length or complexity of certain English records. To maintain consistency, these instances were manually annotated by three annotators, and the emotional tone was determined based on the majority agreement. Finally, each record was broken down into positive, negative, or neutral categories.

The pie charts and bar charts in Figure 5 present the sentiment proportion of all records in English and Chinese. It indicates that over 50% of the information, in both English and Chinese, is characterized as negative. Specifically, English records have a negative proportion of 59.96%, while Chinese records have a negative proportion of 50.71%. This implies that regardless of the language system, over half or more of the infodemic being disseminated to the public carries a negative tone. In addition, the distribution of positive and neutral information differs between the two language systems. Within the Chinese records, there is a balance between positive (25.69%) and neutral (23.60%) information. On the other hand, in the English records, the proportion of positive information exceeds that of neutral information significantly, with 31.12% being positive and only 8.92% being neutral. These findings suggest that individuals within the English language system tend to adopt a more positive attitude when confronted with the infodemic during the COVID-19 pandemic. Conversely, individuals in the Chinese language system lean towards a more neutral and conservative stance.

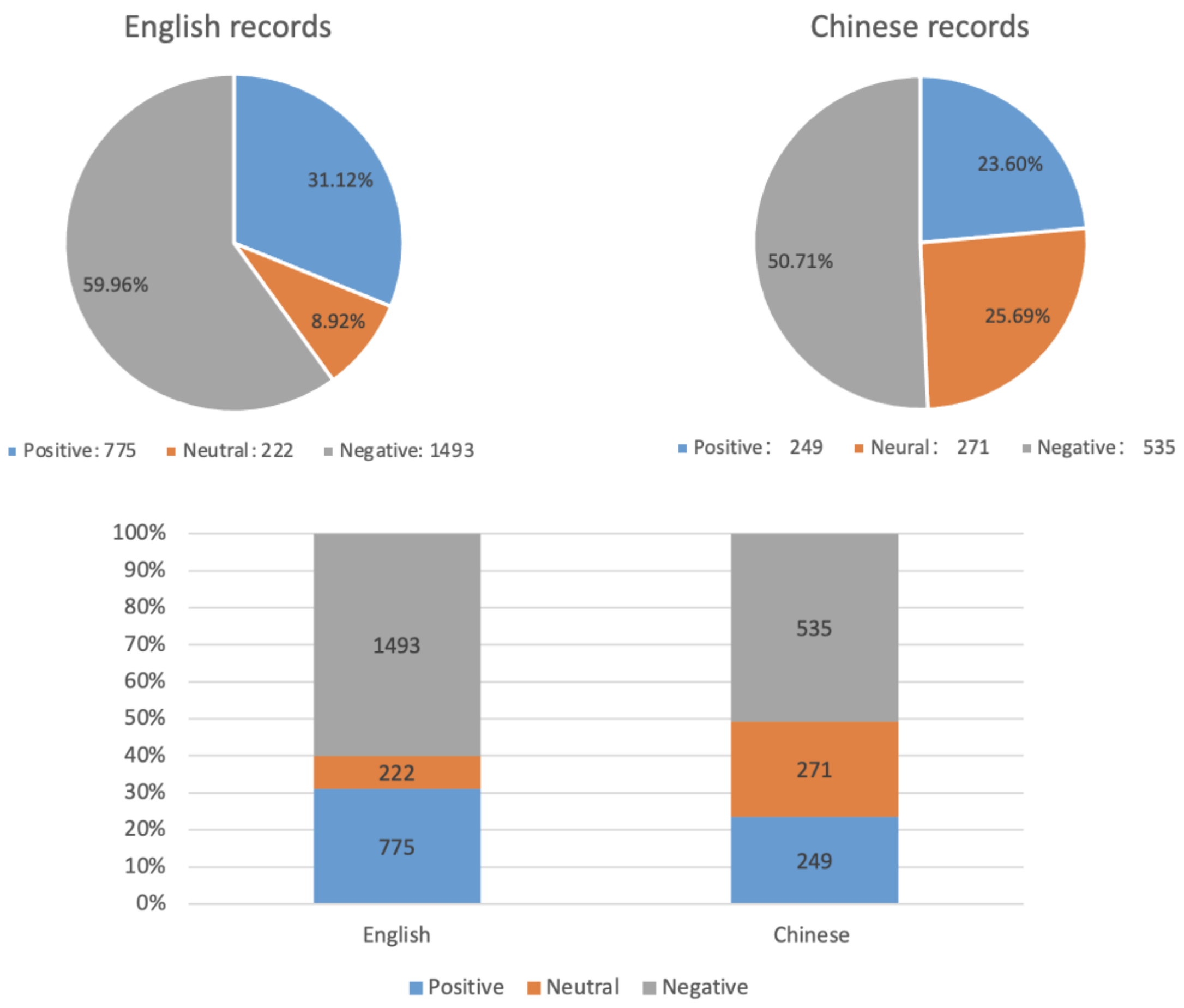

Figure 5. Sentiment analysis results for all records

Pie charts of records labeled as true

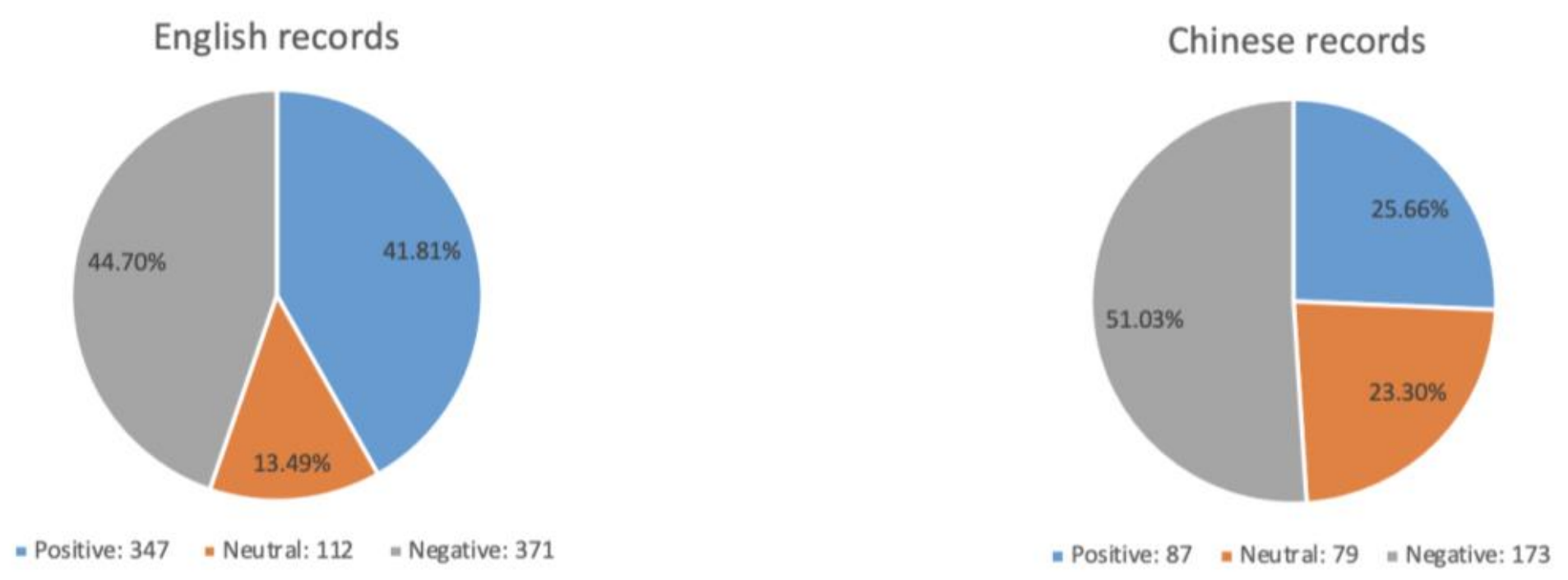

Pie charts of records labeled as false

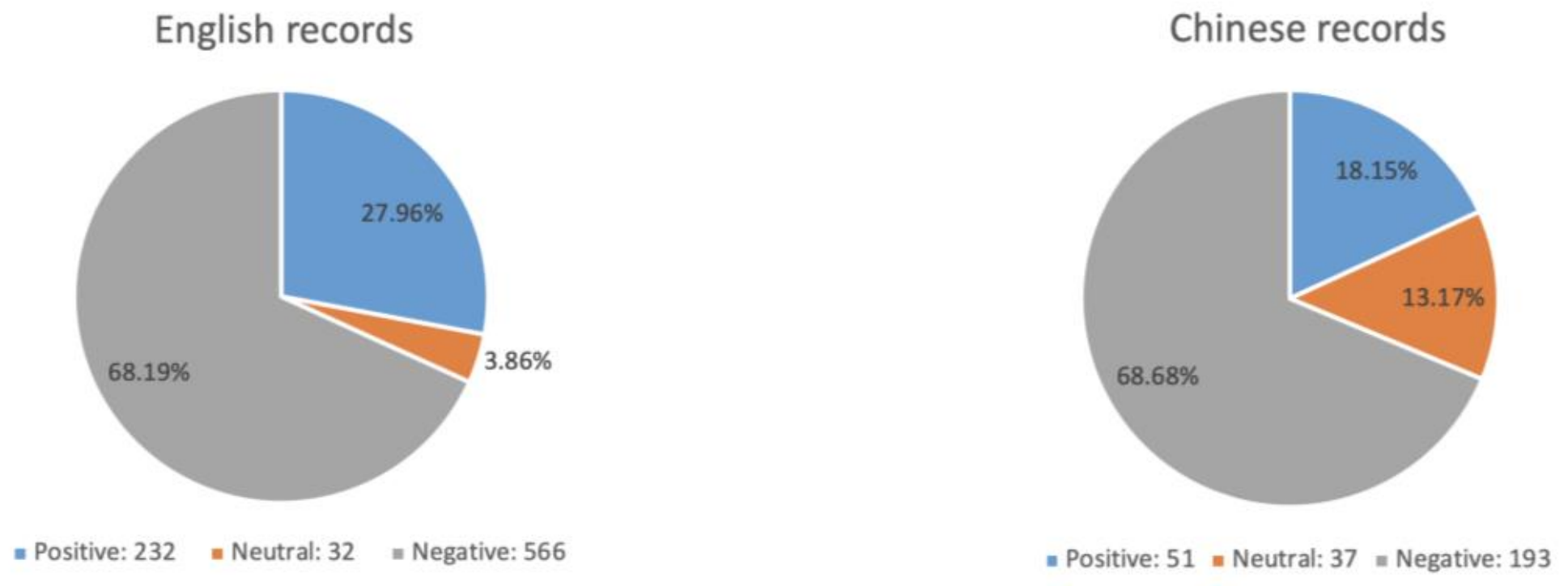

Pie charts of records labeled as uncertain

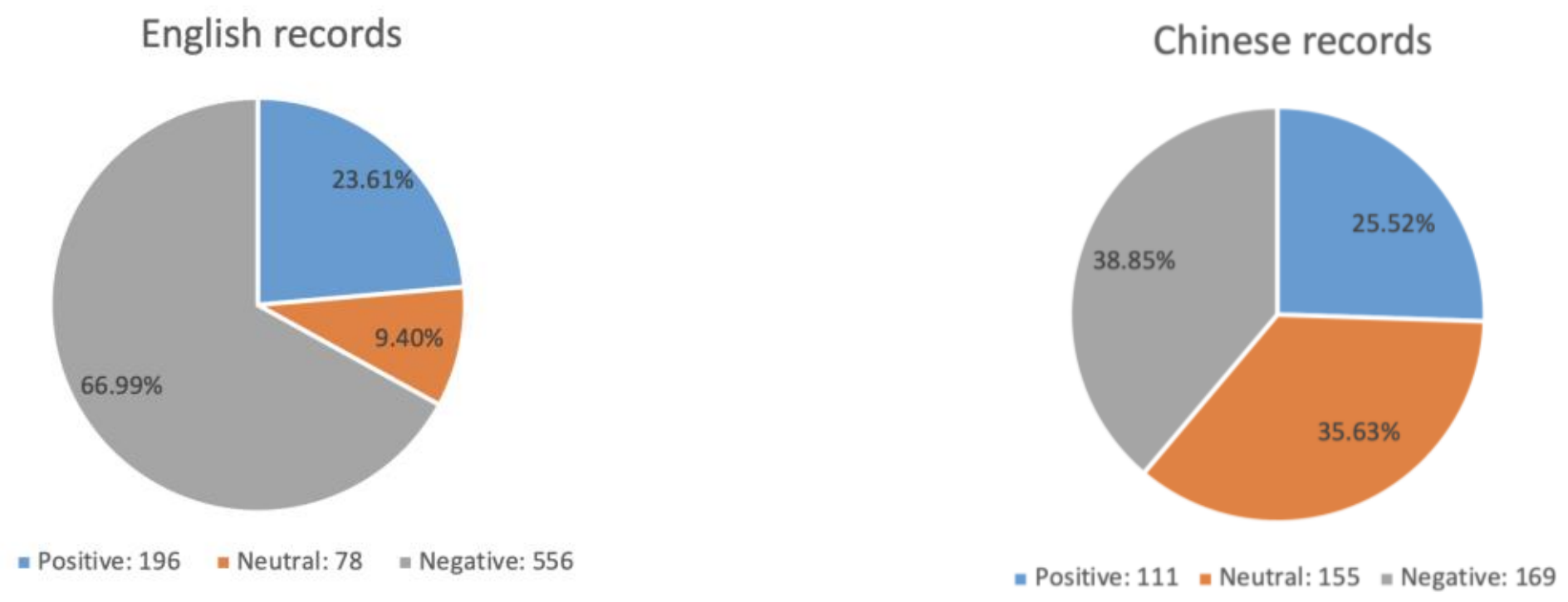

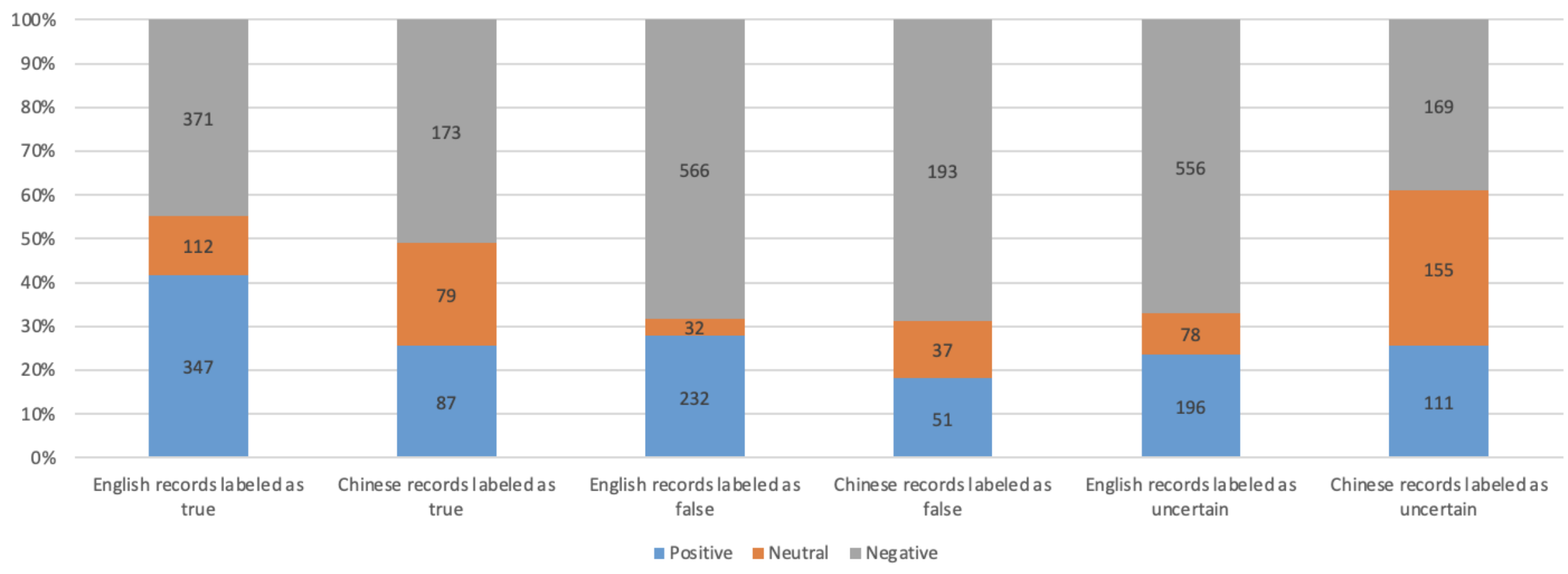

Figure 6. Sentiment analysis results for records classified into three groups

The pie charts and bar charts in Figure 6 categorize records into three groups in both English and Chinese. In the true group, the sentiment proportions for Chinese records have remained relatively stable compared to the results displayed in Figure 5. However, for English records, there has been an increase in the proportions of positive and neutral information. In the false group, it is observed that the proportions of negative information have increased in both languages and remain relatively consistent. Additionally, false records exhibit the highest proportion of negative sentiment among the three groups. Moving on to the uncertain group, the sentiment proportions for English records have not shown significant changes compared to the false group. However, for Chinese records in the uncertain group, the proportion of negative sentiment has decreased, resulting in a relatively balanced distribution of the three sentiment categories.

## 5 Discussion

Regarding word frequency analysis, the distinctions between the English and Chinese word clouds reflect some unique perspectives. Firstly, the term "mask" holds particular significance in the Chinese context, reflecting the country's proactive approach to mask-wearing as a preventive measure against the virus. This cultural aspect is not as prominent in the English word cloud, indicating potential differences in the adoption and perception of this protective measure. Secondly, the variation in the frequency of the term "death" between the English and Chinese word clouds sheds light on the different tones and focuses within each language. The higher occurrence in the English cloud may indicate a greater emphasis on the global loss of life and the severity of the situation, whereas its absence in the Chinese cloud might suggest a more limited or sensitive discussion surrounding this aspect. Thirdly, the individuals most frequently mentioned, President Donald Trump in English and Zhong Nanshan, an esteemed healthcare academician in Chinese, further exemplify the contrasting perspectives. It highlights the significance of political figures in English discussions and the recognition of medical experts and authoritative voices in the Chinese discourse. Finally, the region-specific emphasis in the Chinese cloud and the temporal focus in the English cloud showcase the nuances and contextual factors shaping the discussions in each language. These city names suggest a focus on regional impact and potential localized concerns within China while these time-related terms reflect the need to stay updated with real-time information within English conversations.

The topic clustering analysis highlights the distinct characteristics and priorities within the English and Chinese discussions on COVID-19. Firstly, the English records have a lower number of clustered topics compared to the Chinese records in most cases. This discrepancy suggests that the English discussions on COVID-19 may exhibit a more focused and limited scope while the Chinese records suggest a wider range of perspectives and a more nuanced understanding of various aspects. Secondly, the group labeled as false stands out as an exception to this pattern, with the English records displaying a significantly larger number of clustered topics compared to the Chinese records. This stark difference may indicate a higher prevalence of diverse false narratives and misinformation spread across various sources within the English language. Thirdly, the presence of shared terms in the top 30 relevant terms of Topic 1 signifies a shared focus between both languages, particularly within the group labeled as true and false. This common attention highlights the significance of specific themes or concerns in the global discourse surrounding COVID-19, transcending linguistic and cultural boundaries. Finally, there are noticeable differences in the top 30 relevant terms of Topic 1 within the uncertain group between the Chinese and English languages. These variations emphasize disparities in how uncertain records are conceptualized and discussed within the Chinese and English language communities, which can likely be attributed to variances in cultural, linguistic, and contextual factors.

When it comes to sentiment analysis, the comparative results in English and Chinese records offer valuable insights into emotional trends. Firstly, it is evident that in most cases, over 50% of the information in both languages skews towards negativity, indicating a prevalent negative sentiment in the collected infodemic data. This is likely influenced by the nature of the discussed topics, the tone employed, and the general sentiment of those generating the records. Secondly, English records typically demonstrate a notably higher proportion of positive sentiment with a substantial margin compared to their Chinese counterparts. This disparity can be attributed to various factors, such as cultural contexts, linguistic nuances, or even the diverse user demographics associated with each language. Thirdly, false records consistently manifest the highest proportion of negative sentiment among the three groups in both languages. This observation implies a strong association between misinformation and the generation of negative sentiment among readers. As a result, there is a critical need to actively combat the spread of false records since misleading content not only deceives individuals but also significantly impacts their emotional well-being. Finally, English records in the uncertain group display a nearly identical proportion of negative sentiment, while Chinese records show a decline in negative sentiment. This divergence implies a potential shift towards increased clarity or certainty in the Chinese records classified as uncertain, suggesting that Chinese sources may provide more conclusive or reliable content in this group compared to their English counterparts.

There are several limitations to this study. Firstly, most data collected from the two datasets were obtained from authoritative and representative channels that specifically focus on gathering and presenting valuable information related to popular online topics. However, relying on these sources can introduce biases as the selection of sources and editorial decisions may influence the representation of different perspectives and prioritize certain viewpoints. Secondly, the English records' labels are determined by invited healthcare workers' judgments using a majority agreement methodology. This approach can lead to variations in labeling due to individual differences in interpretation, knowledge, and biases. The absence of clear guidance or standardized criteria for healthcare workers further contributes to potential inconsistencies in labeling decisions. Thirdly, the retention rate for English data sourced from [20] is low. After manual classification, only 830 records were included in the real group, which is the smallest group out of the three. Ultimately, a total of 2490 records were retained for equal distribution among each group. Considering the initial count of 6420 records, the overall retention rate is only 38.78%.

## 6 Conclusions and Future Works

This paper presents a comparative analysis of the COVID-19 infodemic in English and Chinese languages, utilizing textual data extracted from social media platforms. Firstly, to ensure a balanced representation and a fair assessment, two infodemic datasets were introduced through the augmentation of previously collected social media textual data with annotations provided by healthcare workers. Secondly, word frequency analysis was conducted, revealing the thirty-five most frequently occurring infodemic words in both English and Chinese. This comparison offers valuable insights into the prevalent discussions surrounding the COVID-19 infodemic. Thirdly, topic clustering analysis was performed to identify thematic structures present in both languages. This exploration provides a deeper understanding of the primary topics related to the COVID-19 infodemic within each language context. Finally, sentiment analysis was carried out to evaluate the distribution of positive, neutral, and negative sentiments.

This investigation helps comprehend the overall emotional tone associated with COVID-19 information shared on social media platforms in the English and Chinese languages.

In the future, we intend to conduct a study considering the contextual factors. The two proposed datasets in this paper solely consist of original posts from social media, excluding reposts and replies. Additionally, certain records were sourced from official handbooks, authoritative webpages, and fact-verification websites, which lack propagation information. Therefore, the first issue is to collect the user social engagements from the social platform based on infodemic content, including the timestamp of who engages in the records dissemination process. The second line of interest is to conduct a comprehensive understanding of how infodemic spreads within the online community by effectively analyzing users' interactions and their engagement records. Finally, an in-depth analysis will be implemented to seek valuable insights into the mechanisms and dynamics of infodemic propagation, aiming to uncover why and how infodemics occur.

**Author Contributions:**

Jia Luo initiated the idea, addressed whole issues in the manuscript, and wrote the manuscript. Daiyun Peng conducted the numerical experiments. Lei Shi wrote and revised the manuscript. Didier El-baz revised and polished the final edition of the manuscript. Xinran Liu verified the dataset. All authors have read and agreed to the published version of the manuscript.

**Funding:**

This work is supported by the National Natural Science Foundation of China (Grant No. 72104016), the R&D Program of the Beijing Municipal Education Commission (Grant No. SM202110005011), the Natural Science Foundation of Chongqing, China (Grant No. CSTB2023NSCQ-MSX0391), and the Guangxi Key Laboratory of Trusted Software (No. KX202315).

**Data Availability Statement:**

The proposed two datasets are available at https://www.dropbox.com/scl/fo/1qug1snyu49bsiuj53hty/h?rlkey=kew7715ubl83jhmtvcroxv7uj&dl=0 (accessed on 11 August 2023).

**Conflicts of Interest:**

The authors declare no potential conflicts of interest concerning the research, authorship, and/or publication of this article.

**Appendix 1:** Translation for Chinese displayed in Figure 1 and Figure 2

| All records | | Records labeled as true | | Records labeled as false | | Records labeled uncertain | |
|---|---|---|---|---|---|---|---|
| Chinese | Translation | Chinese | Translation | Chinese | Translation | Chinese | Translation |
| 病毒 | Virus | 病毒 | Virus | 病毒 | Virus | 肺炎 | Pneumonia |
| 肺炎 | Pneumonia | 口罩 | Mask | 肺炎 | Pneumonia | 武汉 | Wuhan |
| 口罩 | Mask | 肺炎 | Pneumonia | 口罩 | Mask | 病毒 | Virus |
| 疫情 | Epidemic | 患者 | Patient | 疫情 | Epidemic | 疫情 | Epidemic |
| 武汉 | Wuhan | 消毒剂 | Sanitizer | 患者 | Patient | 医院 | Hospital |
| 患者 | Patient | 症状 | Symptom | 美国 | U.S. | 美国 | U.S. |
| 美国 | U.S. | 医用 | Medical | 酒精 | Ethyl alcohol | 中国 | China |
| 钟南山 | Zhong Nanshan | 飞沫 | Droplet Infection | 钟南山 | Zhong Nanshan | 口罩 | Mask |
| 消毒剂 | Sanitizer | 建议 | Suggestion | 疫苗 | Vaccine | 钟南山 | Zhong Nanshan |
| 酒精 | Ethyl alcohol | 风险 | Risk | 武汉 | Wuhan | 患者 | Patient |
| 疫苗 | Vaccine | 酒精 | Ethyl alcohol | 大蒜 | Garlic | 北京 | Beijing |
| 医院 | Hospital | 疾病 | Disease | 大量 | Abundant | 上海 | Shanghai |
| 中国 | China | 证据 | Evidence | 病人 | Patient | 意大利 | Italy |
| 病人 | Patient | 感染者 | Infected person | 日本 | Japan | 病人 | Patient |
| 症状 | Symptom | 人群 | Crowd | 抗体 | Antibody | 疫苗 | Vaccine |
| 医用 | Medical | 儿童 | Children | 院士 | Academician | 病例 | Patient case |
| 风险 | Risk | 居家 | Staying at home | 医生 | Doctor | 湖北 | Hubei province |
| 北京 | Beijing | 通风 | Ventilating | 病毒感染 | Virus infection | 人员 | Staff |
| 人员 | Staff | 人员 | Staff | 空气 | Air | 成都 | Chengdu |
| 病例 | Patient case | 物品 | Goods | 白酒 | Liquor | 院士 | Academician |
| 意大利 | Italy | 效果 | Effect | 防病毒 | Anti-virus | 入境 | Immigration |
| 建议 | Suggestion | 传染性 | Infectiousness | 小时 | Hours | 医生 | Doctor |
| 上海 | Shanghai | 人类 | Human | 病情 | Illness state | 视频 | Video |
| 飞沫 | Droplet Infection | 距离 | Distance | 中国 | China | 全国 | Nationwide |
| 抗体 | Antibody | 核酸检测 | PCR test | 流鼻涕 | Rhinorrhea | 全部 | Entire |
| 院士 | Academician | 疫苗 | Vaccine | 纸尿裤 | Diaper | 阳性 | Positive |
| 感染者 | Infected person | 动物 | Animal | 气溶胶 | Aerosol | 员工 | Staff |
| 阳性 | Positive | 食品 | Food | 二氧化氯 | Chlorine dioxide | 印度 | India |
| 核酸检测 | PCR test | 情况 | Situation | 消毒剂 | Sanitizer | 国家 | Country |
| 医生 | Doctor | 传播者 | Spreader | 牛羊肉 | Beef and mutton | 物资 | Goods |
| 疾病 | Disease | 重症 | Severe case | 喉咙 | Throat | 酒精 | Ethyl alcohol |
| 湖北 | Hubei province | 手部 | Hand | 肥皂 | Soap | 特朗普 | Trump |
| 空气 | Air | 手套 | Glove | 食品 | Food | 风险 | Risk |
| 证据 | Evidence | 传染病 | Infectious disease | 食用 | Edible | 广州 | Canton |
| 人类 | Human | 紫外线 | Ultraviolet ray | 瘟疫 | Plague | 医疗 | Medical treatment |

**Appendix 2:** Translation for Chinese displayed in Figure 3 and Figure 4

| All records | | Records labeled as true | | Records labeled as false | | Records labeled uncertain | |
|---|---|---|---|---|---|---|---|
| Chinese | Translation | Chinese | Translation | Chinese | Translation | Chinese | Translation |
| 新冠 | Covid-19 | 新冠 | Covid-19 | 新型 | Novel | 新冠 | Covid-19 |
| 病毒 | Virus | 病毒 | Virus | 预防 | Prevention | 感染 | Infection |
| 冠状病毒 | Coronavirus | 感染 | Infection | 冠状病毒 | Coronavirus | 武汉 | Wuhan |
| 肺炎 | Pneumonia | 肺炎 | Pneumonia | 病毒 | Virus | 肺炎 | Pneumonia |
| 新型 | Novel | 冠状病毒 | Coronavirus | 新冠 | Covid-19 | 病毒 | Virus |
| 感染 | Infection | 时 | Hours, during | 肺炎 | Pneumonia | 月 | Months |
| 口罩 | Mask | 洗手 | Washing hands | 消毒 | Sterilizing | 天 | Days |
| 预防 | Prevention | 没有 | No, without | 感染 | Infection | 美国 | U.S. |
| 没有 | No, without | 消毒剂 | Sanitizer | 口罩 | Mask | 口罩 | Mask |
| 传播 | Spreading | 接触 | Touching | 酒精 | Ethyl alcohol | 成都 | Chengdu |
| 可能 | Maybe | 口罩 | Mask | 治疗 | Curing | 广州 | Guangzhou |
| 使用 | Using | 戴 | Wearing | 有效 | Effective | 高速 | High way |
| 隔离 | Quarantine | 检测 | Test | 30 | 30 | 疫苗 | Vaccine |
| 会 | Able | 患者 | Patient | 秒 | Seconds | 会 | Able |
| 不能 | Unable | 传播 | Spreading | 传播 | Spreading | 日 | Date |
| 接触 | Touching | 可能 | Maybe | 75% | 75% | 病人 | Patient |
| 治疗 | Curing | 含氯 | Chlorinated | 美国 | U.S. | 养老院 | Retirement home |
| 医院 | Hospital | 治疗 | Curing | 武汉 | Wuhan | 消毒 | Sterilizing |
| 防护 | Protection | 有效 | Effective | 疫情 | Epidemic | 中国 | China |
| 疫苗 | Vaccine | 使用 | Using | 喝 | Drinking | 确诊 | Confirming |
| 时 | Hours, during | 疫苗 | Vaccine | 医生 | Doctor | 抗体 | Antibody |
| 有效 | Effective | 疾病 | Disease | 钟南山 | Zhong Nanshan | 日晚 | Evening |
| 避免 | Avoiding | 呼吸道 | Respiratory tract | 度 | Degree | 发现 | Discovering |
| 武汉 | Wuhan | 预防 | Prevention | 容易 | Easy | 上海 | Shanghai |
| 清洁 | Clean | 不能 | Unable | 中国 | China | 隔离 | Quarantine |
| C | C | 新型 | Novel | 喉咙 | Throat | 加拿大 | Canada |
| 佩戴 | Wearing | 应 | Should | 日本 | Japan | 京东 | JD.com |
| 中 | In, China | 中 | In, China | SARS | SARS | 23 | 23 |
| 容易 | Easy | 核酸 | Pcr | 开水 | Boiled water | 增加 | Increasing |
| 人类 | Human | 天 | Days | 戴 | Wearing | 宾馆 | Hotel |

**Reference:**

[1]. Zarocostas, J. (2020). How to fight an infodemic. The lancet, 395(10225), 676.
[2]. Xu, J., & Liu, C. (2021). Infodemic vs. pandemic factors associated to public anxiety in the early stage of the COVID-19 outbreak: a cross-sectional study in China. Frontiers in Public Health, 9, 723648.
[3]. www.statista.com/statistics/262946/share-of-the-most-common-languages-on-the-internet/
[4]. Sanaullah, A. R., Das, A., Das, A., Kabir, M. A., & Shu, K. (2022). Applications of machine learning for COVID-19 misinformation: a systematic review. Social Network Analysis and Mining, 12(1), 94.
[5]. Glazkova, A., Glazkov, M., & Trifonov, T. (2021, February). g2tmn at constraint@ aaai2021: exploiting CT-BERT and ensembling learning for COVID-19 fake news detection. In International Workshop on Combating Online Hostile Posts in Regional Languages during Emergency Situation (pp. 116-127). Cham: Springer International Publishing.
[6]. Chen, B., Chen, B., Gao, D., Chen, Q., Huo, C., Meng, X., ... & Zhou, Y. (2021). Transformer-based language model fine-tuning methods for COVID-19 fake news detection. In Combating Online Hostile Posts in Regional Languages during Emergency Situation: First International Workshop, CONSTRAINT 2021, Collocated with AAAI 2021, Virtual Event, February 8, 2021, Revised Selected Papers 1 (pp. 83-92). Springer International Publishing.
[7]. Paka, W. S., Bansal, R., Kaushik, A., Sengupta, S., & Chakraborty, T. (2021). Cross-SEAN: A cross-stitch semi-supervised neural attention model for COVID-19 fake news detection. Applied Soft Computing, 107, 107393.
[8]. Chen, M. Y., & Lai, Y. W. (2022). Using fuzzy clustering with deep learning models for detection of COVID-19 disinformation. Transactions on Asian and Low-Resource Language Information Processing.
[9]. Liu, J., & Chen, M. (2023, February). COVID-19 Fake News Detector. In 2023 International Conference on Computing, Networking and Communications (ICNC) (pp. 463-467). IEEE.
[10]. Gupta, A., Li, H., Farnoush, A., & Jiang, W. (2022). Understanding patterns of COVID infodemic: A systematic and pragmatic approach to curb fake news. Journal of business research, 140, 670-683.
[11]. Wan, M., Su, Q., Xiang, R., & Huang, C. R. (2023). Data-driven analytics of COVID-19 'infodemic'. International journal of data science and analytics, 15(3), 313-327.

[12]. Zhao, J., Fu, C., & Kang, X. (2022). Content characteristics predict the putative authenticity of COVID-19 rumors. Frontiers in Public Health, 10, 920103.
[13]. Zhou, L., Tao, J., & Zhang, D. (2023). Does fake news in different languages tell the same story? An analysis of multi-level thematic and emotional characteristics of news about COVID-19. Information Systems Frontiers, 25(2), 493-512.
[14]. Murayama, T. (2021). Dataset of fake news detection and fact verification: a survey. arXiv preprint arXiv:2111.03299.
[15]. Cheng, M., Wang, S., Yan, X., Yang, T., Wang, W., Huang, Z., ... & Bogdan, P. (2021). A COVID-19 rumor dataset. Frontiers in Psychology, 12, 644801.
[16].Haouari, F., Hasanain, M., Suwaileh, R., & Elsayed, T. (2020). ArCOV19-rumors: Arabic COVID-19 twitter dataset for misinformation detection. arXiv preprint arXiv:2010.08768.
[17]. Luo, J., Xue, R., Hu, J., & El Baz, D. (2021). Combating the Infodemic: A Chinese Infodemic Dataset for Misinformation Identification. Healthcare, 9(9), 1094.
[18]. Kim, J., Aum, J., Lee, S., Jang, Y., Park, E., & Choi, D. (2021). FibVID: Comprehensive fake news diffusion dataset during the COVID-19 period. Telematics and Informatics, 64, 101688.
[19]. Dharawat, A. R., Lourentzou, I., Morales, A., & Zhai, C. (2020). Drink bleach or do what now? covid-hera: A dataset for risk-informed health decision making in the presence of covid19 misinformation.
[20]. Patwa, P., Sharma, S., Pykl, S., Guptha, V., Kumari, G., Akhtar, M. S., ... & Chakraborty, T. (2021). Fighting an infodemic: Covid-19 fake news dataset. In Combating Online Hostile Posts in Regional Languages during Emergency Situation: First International Workshop, CONSTRAINT 2021, Collocated with AAAI 2021, Virtual Event, February 8, 2021, Revised Selected Papers 1 (pp. 21-29). Springer International Publishing.
[21]. https://constraint-shared-task-2021.github.io/
[22]. https://www.weiciyun.com/
[23]. Blei, D. M., Ng, A. Y., & Jordan, M. I. (2003). Latent dirichlet allocation. Journal of machine Learning research, 3(Jan), 993-1022.
[24]. Sievert, C., Shirley, K., & Davis, L. A method for visualizing and interpreting topics. In Proceedings of Workshop on Interactive Language Learning, Visualization, and Interfaces, Association for Computational Linguistics (pp. 63-70).
[25]. https://monkeylearn.com
[26]. Zhang, B., Jiang, Y., & Zhou, J. (2021). Analysis of the Contents of the “Draft of the Preschool Education Law of the People’s Republic Of China (Draft for Solicitation of Comments)” Based on the ROST CM6. 0 Content Mining System. Chinese Education & Society, 54(1-2), 1-20.